# Expanding photonic palette: exploring high index materials.

Jacob B Khurgin
Department of Electrical and Computer Engineering
Johns Hopkins University
Baltimore MD 21218

**ABSTRACT:** While the photonic community is being occupied with exotic concepts portending a grand future and fame if not a fortune, I respectfully entertain the possibility that a humble concept of simply increasing refractive index by a modest factor may have a far greater payoff in many walks of life. With that in mind, I explore why higher index materials have not yet materialized, and point out a few tentative directions for the search of these elusive materials, be they natural or artificial.

- **INTRODUCTION**

In every technology, material properties play a paramount role, and among those properties, one can pinpoint a few that ultimately enable a wide variety of devices and techniques. For electronics, such key material properties are conductivity, carrier mobility, and dielectric constant; for thermal control, it is thermal conductivity and specific heat; for micromechanics, it is Young's modulus, hardness, and elasticity; for acoustics, it is the speed of sound, and so on. What is common among all of these properties is that Nature and human determination have given us an extremely wide range of materials where the aforementioned (and many other) properties vary by many orders of magnitude. This provides engineers and designers with a very broad design space that is continuously expanding as new materials enter mainstream. However, the situation is quite different in photonics where the defining material property is the refractive index. Here, the refractive index $n$ determines the minimum achievable features in imaging and lithography through the diffraction limit, $\lambda/n$, while the index contrast between two materials determines the minimum size of integrated optical components and optical strength of components, such as lenses, diffraction elements, and metasurfaces.

Although the index can be as high as 3.5 in near IR (Si) and 5.8 in mid-IR (PbTe), given its importance, it is astonishing that since the days of Carl Zeiss the range of refractive indices in high-quality optical materials widely available in the visible range has been limited to 1.38 (MgF$_2$) to 2.4 (TiO$_2$) as one can see from Fig.1[1]

It would be expected that with all the revolutionary developments in materials science that took hold in the last decades, a substantial effort would be directed at expanding the limited palette of refractive indices available to photonics practitioners. Yet, clearly it does not appear to be the case. Instead, through the last couple of decades' photonics community has been on wild goose chase of far more exotic materials, with negative[2-3] or near-zero[4] index, topological materials[5], and ubiquitous 2D materials[6], graphene[7] being the king of them all. A true deluge of titles in reputable publications has materialized, loaded with superlatives like "Giant", "Super", "Colossal", "Ultra", Extraordinary", "Exceptional", "Perfect", and…the reader can continue the list *ad infinitum*. The pay-off so far has been rather modest, with no practical low loss negative and near-zero index materials emerging, and topological materials (in my view) being more of a solution in search of a problem. As far as two dimensional materials are concerned, while they may exhibit interesting effects per unit thickness, all their purported advantages happen to be in vain, as more than a single atomic layer is required for any optical phenomenon (other than absorption) to have appreciable efficiency [8], sufficient to be presented on a graph with a vertical scale featuring some meaningful units rather than ubiquitous "*arb. u*". One can also add to the above list the infatuation with metals (plasmonics[9]) even when faced with the undeniable fact that unlike electronics, where amplification is easily available, photonics is far less tolerant of the loss, and the loss in metals is inherently huge[10]. Hence plasmonics has not found too many applications beyond sensing[11] and newly emerging science of single photon emission[12] (whose ultimate value hinges upon eventual spread of quantum information processing, a topic on which I do not dare to speculate as regrettably I do not expect to witness it in my lifetime[13]). Least it creates an impression of me unjustly criticizing the community, here is *mea culpa*: I have been just as guilty of participating in mad scramble for the next fashionable material *du jour* as anyone else, publishing extensively in many of the aforementioned fields, although perhaps trying to retain some vestiges of sober judgment and not get carried away with my own rhetoric. Obviously, I am not the first one to raise these questions and many honest and fair (and hence unpopular) critical assessments of recently or currently "hot" topics can be found in the literature[13-18].

Moving from this constructive criticism to a more positive discussion, it is hard to avoid the topic of metamaterials[19], a term, which these days encompasses just about anything that can be conceivably fabricated (or envisioned) with diverse small scale techniques, be it lithography, nanoassembly, or epitaxial growth. While the 3-dimensional metamaterials have not lived up to their early promise, their 2-dimensional offspring, metasurfaces[20], have withstood the reality test and already are rapidly becoming an integral part of photonic techniques and devices. The original ideas for metasurfaces all relied on using metals[21], but the metal metastructures have not advanced far beyond the publication in reputable journal stage (not to say that this stage is not critically important in assuring uninterrupted supply of scientific manpower and funding). The reasons are multiple, but one can relate them all to a hardly groundbreaking declaration – *just as it is problematic to cook up a delicious dish from bad ingredients, it is also hard to cobble together a high quality optical metamaterial (surface) from low optical quality materials*, such as metals with their high loss. At the same time, all-dielectric metasurfaces[22-23], using true and tried high quality and low loss materials did take off and have shown performance metrics that are equal or even surpassing those of conventional optics[23-24]. But note that a large, and perhaps defining factor in metasurfaces success has been incorporation of high index material – $TiO_2$ with index of 2.4 in contrast to the conventional optics relying on $SiO_2$ (n=1.45). With optical power proportional to $n-1$ it is no wonder that meta lenses outperform silica optics by a factor of 3!. Similarly, in near IR region, Si (n=3.4) on $SiO_2$ metasurfaces exhibit enviable performance characteristics. But even with these highest available indices, most metasurfaces fabricated today do not look at all like a flat prairie landscape as it was envisioned at their conception. In fact, under an electronic microscope, they look more like the island of Manhattan with tall and narrow pillars of skyscrapers arranged in the square blocks, high aspect ratios[25] required to attain phase and group delays needed to achieve decent performance. In the absence of higher index materials one is forced to further increase the height of metasurfaces creating fabrication difficulties.

Therefore, it is not really a stretch to imagine that if even higher refractive indices were available to the designers, the performance of metasurfaces would reach heights that today appear unimaginable.

Through the years, the argument for discovering high index materials has been rather narrowly focused on one obvious goal - with higher index one can increase resolution of imaging optics, reduce weight and thickness of lenses and also facilitate aberration correction. Now, however, the advent of metasurfaces, photonic crystals and photonic integrated circuits (PICs) makes the quest for high index materials if anything more urgent and important as performance in all these techniques is hinged upon high index contrast. With materials having higher refractive index (while obviously maintaining transparency), one can envision PICs with smaller footprint devices, photonic crystals that exhibit wide bandgaps in 3 dimensions, large numerical aperture metasurfaces, high resolution immersion optics, ultra-narrow waveguides, and many other devices. I would even venture to mention how high index materials may benefit such esoteric concepts as optical cloaking and transformational optics, but prefer not to be carried away, leaving these enticing topics to more daring souls, and concentrate on earthlier concepts.

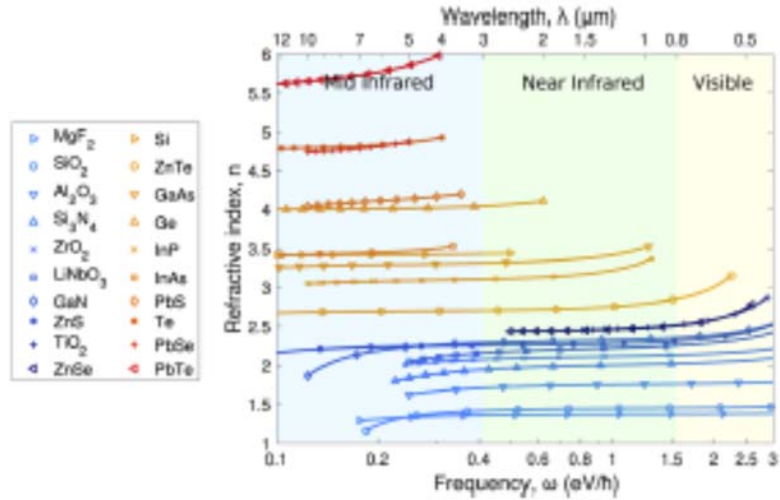

**Figure 1** Indices of refraction for widely available materials [1]

To summarize, perhaps the time has come for us to take a fresh look at the issue that has been considered settled around 19th century, if not earlier, namely: why are the values of refractive indices what they are, and what is that prevents us from getting the larger values? In this essay, I embark on this less trodden path. First I remind the readers about the origin of refractive index in condensed matter, emphasizing the fact that unlike absorption properties that are determined mostly by the states near the lowest bandgap, the refraction is determined by the states throughout the entire Brillouin zone. It makes estimate both more difficult and at the same time easy, if one makes a few realistic approximations. Then a simple polarizable bond model[26] will be reviewed, the estimate of the refractive index far will be made, and the limitations will naturally emerge. Following that the different means to achieve high index combine with low loss in natural and artificial materials will be considered with a guarded optimism expressed, stating that 50-100% enhancement of refractive index across a relatively narrow band is not entirely out of reach if concerted effort is mounted. While not meriting such aforementioned superlatives as "giant" and "ultra-broad band", the prospects of higher density PICs, more powerful and versatile metalenses, and factor of 2 improvement of resolution in immersion optics should hopefully make this paper worthy of reading beyond this point.

- **WHY IS REFRACTIVE INDEX RANGE SO LIMITED?**

Although it is universally known that the refractive indices range is so small, the reason why is it so is perhaps less well recognized. On the most basic level it follows directly from the classical Lorentz model in which a polarizable electron cloud is kept in its place by Coulomb attraction to one (as in case of isolated atoms) or two or more in case polarizable bonds in molecules and condensed matter. In the most simple textbook example the electron cloud is spherical with the "atomic" radius $r_a$ as shown in Fig.2a.

From the trivial electrostatic considerations the restoring force acting on electron that is moved by distance $\mathbf{r}$ from equilibrium position is the same as the one acting on the ion

$$\mathbf{F}_a = -\frac{e^2}{4\pi\varepsilon_0 r_a^3}\mathbf{r} \equiv -K\mathbf{r}, \qquad (1)$$

where $K$ can be though as a "spring coefficient". The resonant frequency of the Lorentz oscillator is $\omega_0 = \sqrt{K/m_0}$, and the dielectric constant can be found (neglecting damping) as

$$\varepsilon_r = 1 + \chi_{ion} + \frac{\omega_p^2}{\omega_0^2 - \omega^2}, \qquad (2)$$

where $\chi_{ion}$ is ionic susceptibility, while the plasma frequency is $\omega_p = \sqrt{Ne^2/\varepsilon_0 m_0}$, and $N$ is the density of the polarizable electrons. In the range of frequencies that is well below $\omega_0$ (i.e. in the transparency region) but ay above the resonant frequencies of the ionic vibrations (which are typically less than 10 THz) one can approximate $\varepsilon_r \sim 1 - \omega_p^2/\omega_0^2$. Substituting $\omega_0$ and $\omega_p$ we immediately obtain

$$\varepsilon_r(0) \approx 1 + \frac{Ne^2}{m_0\varepsilon_0}\left(\frac{e^2}{4\pi\varepsilon_0 r_a^3 m_0}\right)^{-1} = 1 + 4\pi N r_a^3 \sim 1 + 3f_e, \qquad (3)$$

where $f_e = \frac{4}{3}\pi r_a^3 N$ is a fraction of the volume occupied by the electron cloud. Considering that according to Pauli's principle two electrons with opposite spins can occupy the same space one can assume that $0 < f_e < 2$ which immediately presents us with the range of possible refractive indices between 1 and 2.6, and that is right where the experimental values in the visible range are. Obviously, in the materials that are transparent only in the IR range the refractive index increases to higher value, which may be construed as the increased overlap of the electron associated with material becoming more metallic. In general, there exists an empirical Moss Rule [27] connecting the off-resonant refractive index and energy gap

$$n^4 E_G \approx 95 eV \qquad (4)$$

Although a relative primitive, the picture presented so far very well describes the limits on the refractive indexes imposed by nature. Indeed, the authors of a remarkable recent work [28] have performed an exhaustive search of the data on the refractive index and its dispersion $dn/d\omega$ that confirmed (3). Furthermore, they have shown that since in agreement with (2) the only feasible way of enhancing refractive index is to come close to the resonance, a strict constraint is placed on the refractive index and its dispersion,

$$n \leq \left( \frac{\omega_p^2}{\omega} \frac{dn}{d\omega} \right)^{1/3} \qquad (5)$$

So, the possibility of getting broadband large refractive index appears to be quite remote, but for many modern application broadband operation is not necessary. Furthermore, large chromatic dispersion often comes handy for aberration corrections, especially in hybrid lens/metasurface systems. Then, for a moment keeping a Lorentzian character of (2) and assuming the FWHM of absorption $\Delta\omega$, we can obtain maximum permittivity of

$$\varepsilon_r(\omega_0 - \Delta\omega/2) \approx 1 + \chi_{ion} + \frac{\omega_p^2}{\omega_0^2} Q, \qquad (6)$$

where $Q = \omega_0 / \Delta\omega$. Therefore the resonant refractive index $n_{res} \approx n(0) Q^{1/2}$ can be enhanced by a factor of two-to-three assuming that $Q$ can be as high as 10, and visible refractive index of 5 -7 would have a transformative impact in photonics. On its surface $Q \geq 10$ looks quite realistic, indeed even much narrower optical transitions are routinely observed in gases and in materials doped with, for instance, rare earth ions. But once density of active atoms/ions/molecules increases, which is necessary for high index, harsh reality sets in as discrete energy levels broaden into energy bands, to the degree that for most solid materials $Q$ only slightly exceeds unity – hence the resonant enhancement of refractive index is at best 30-40% for ionic materials like LiF, where the refractive index is small anyway and far less for high index covalent materials, like diamond where the resonant enhancement is only about 10%.

To understand how the broadening affects the refractive index a more refined approach is needed. For that we shall revert to the bond theory of dielectric response [26, 29-33], developed more than a half century ago, i.e. when all the towering figures of that period have been busy producing hype-free gems of papers in Physical Review, as glossier publications were yet to be born, and the term "high impact" was primarily invoked in the context of lethal or near-lethal car wrecks.

- **A MORE REFINED LOOK AT THE PROBLEM**

Leaving this nostalgic digression aside, according to the aforementioned works the dielectric constant in the optical range is related to bond polarizabilities in the material. The occupied (unoccupied) bond states $u_v(\mathbf{r})$ ( $u_c(\mathbf{r})$ ) have bonding (anti-bonding) character and correspond to valence (conduction bands). Bond polarizabilities are defined by the transition dipole $e\mathbf{d}_{vc} = e\langle u_v|\mathbf{r}|u_c\rangle$ as shown in Fig. 2b for the case of or zinc blende or wurtzite lattice characteristic of many covalent materials such as II-VI, III-V, as well as group IV (in which case it is called diamond lattice). In each case there four hybrid orbitals in the tetrahedral arrangement surrounding each atom as shown in Fig.2.

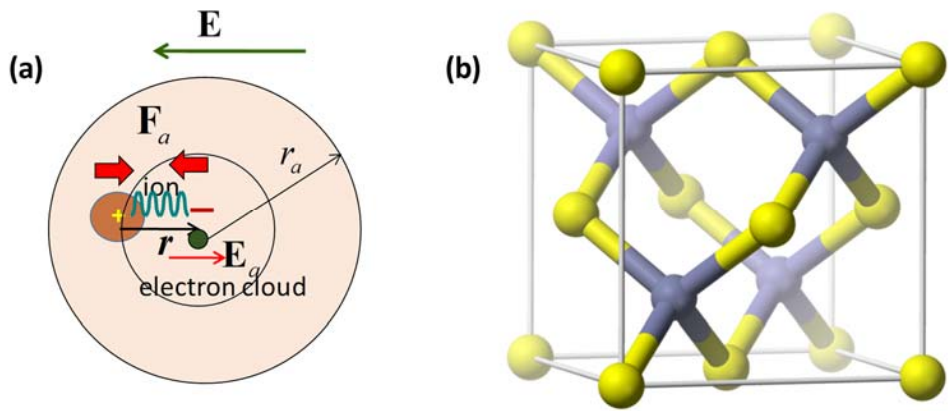

**Figure 2.** (a) Lorentz model used in classical derivation of dielectric permittivity. **E** is external field and $\mathbf{E}_a$ is the intrinsic field of electron cloud.( b) Tetrahedral bonds in a diamond or zinc blend lattice

Summing over all the bonds one can obtain the expression for the dielectric constant

$$\varepsilon_r = 1 + \chi_{ion} + \frac{2E_{vc} N e^2 d_{cv}^2 / \varepsilon_0}{E_{vc}^2 - (\hbar\omega)^2}, \qquad (7)$$

where $E_{vc}$ is the transition energy the two states. Introducing the oscillator strength as

$$f_{cv} = 2m_0 E_{vc} d_{cv}^2 / \hbar^2 \qquad (8)$$

we easily obtain the same equation (2) modified by the oscillator strength,

$$\varepsilon_r = 1 + \chi_{ion} + \frac{f_{cv}\omega_p^2}{\omega_{cv}^2 - \omega^2}. \qquad (9)$$

For the strong allowed transition $f_{cv}$ approaches unity and essentially the same limitations to refractive index do apply.

The bonds, however, are broadened into the bands, each state being a function of wave-vector $\mathbf{k}$ and written as $\Psi_n(\mathbf{k}) = u_n(\mathbf{k})\exp(i\mathbf{k}\cdot\mathbf{r})$, where $u_n(\mathbf{k})$ is the periodic Bloch function and index n stands for conduction (c) or valence (v) band. Therefore, the dielectric constant has to be evaluated as a sum over all the possible transitions between states in two bands

$$\varepsilon_r(\omega) = 1 + \chi_{ion} + \frac{e^2}{\varepsilon_0 V}\sum_{\mathbf{k}}\frac{2E_{vc}(\mathbf{k})d_{cv}^2(\mathbf{k})}{E_{vc}^2(\mathbf{k}) - (\hbar\omega)^2}, \qquad (10)$$

where $V$ is the volume, and also include imaginary part (neglecting off-resonant absorption by ions)

$$\varepsilon_i(\omega) = \frac{\pi e^2}{\varepsilon_0 V}\sum_{\mathbf{k}} d_{cv}^2(\mathbf{k})\delta\left(E_{vc}(\mathbf{k}) - \hbar\omega\right), \qquad (11)$$

Strictly speaking, the dipole moment is not easily defined in the extended periodic system and for calculating of the resonant phenomena, such as absorption spectrum (11) one typically uses matrix element of momentum[34], $\mathbf{p}_{vc} = i\hbar\langle u_v | \nabla | u_c \rangle$, however, if one introduces $\mathbf{d}_{vc}(\mathbf{k}) = \mathbf{p}_{vc}(\mathbf{k})/m\omega_{vc}(\mathbf{k})$, for the calculations of the off-resonant phenomena, such as refractive index the co-ordinate rather than momentum gauge becomes preferable, because one can truncate the summation in (10) at high energies[35-36]

Let us now take a look at characteristic band structures of Si and GaAs shown in Fig.3a and b respectively. At first glance the picture looks messy and bears no resemblance to the bond orbital picture of Fig.2b. At closer look, one realizes that in 3D Brillouin zone (BZ) most of the states are located close to the zone edges, i.e. $X$ and $L$ points. The states near the conduction and valence band extrema, separated by the bandgap energy $E_{gap}$ may play the central role in determining absorption and emission properties (as well as electronic properties in doped materials), but their contribution to the dielectric constant and refractive index is negligibly small, simply because there are so few of them. That is why the refractive indices of direct (GaAs) and indirect (Si) semiconductors are remarkably similar in the transparency region as shown in Fig.3 c and d where the spectra of real and imaginary parts of permittivity are plotted.

One can also notice that along directions pointing away from the center of BZ, the CB and VB run parallel to each other[26] as highlighted in Fig.3 and b. Therefore, large number of transitions have the energy close to $E_1$ and $E_2$, and the spectrum of $\varepsilon_i$, i.e. absorption, shows distinct peaks at those energies, and, accordingly the real part of $\varepsilon$ experiences a Lorentz-like dispersion in the vicinity of these frequencies, becoming negative past $E_2$. Clearly these resonances are nothing but the original resonances associated with tetrahedral hybrid orbitals broadened by the coupling between the individual orbitals. And this fact, of course, presents one with an opportunity to represent the dielectric constant (at least in the transparency region) well above the phonon resonances as

$$\varepsilon_r \approx 1 + \frac{\hbar^2 \omega_p^2}{E_P^2 - \hbar^2 \omega^2}, \tag{12}$$

where $E_P$ is called the Penn gap[32] that is typically in between $E_1$ and $E_2$, closer to $E_2$.[37] For example, for Si $E_P \approx 4.8 eV$ and for GaAs $E_P \approx 5.1 eV$. One may think of Penn gap as some average transition energy in which most of the oscillator strength is concentrated and which defines the refractive index, while the fundamental bandgap $E_G$ defines the absorption properties.

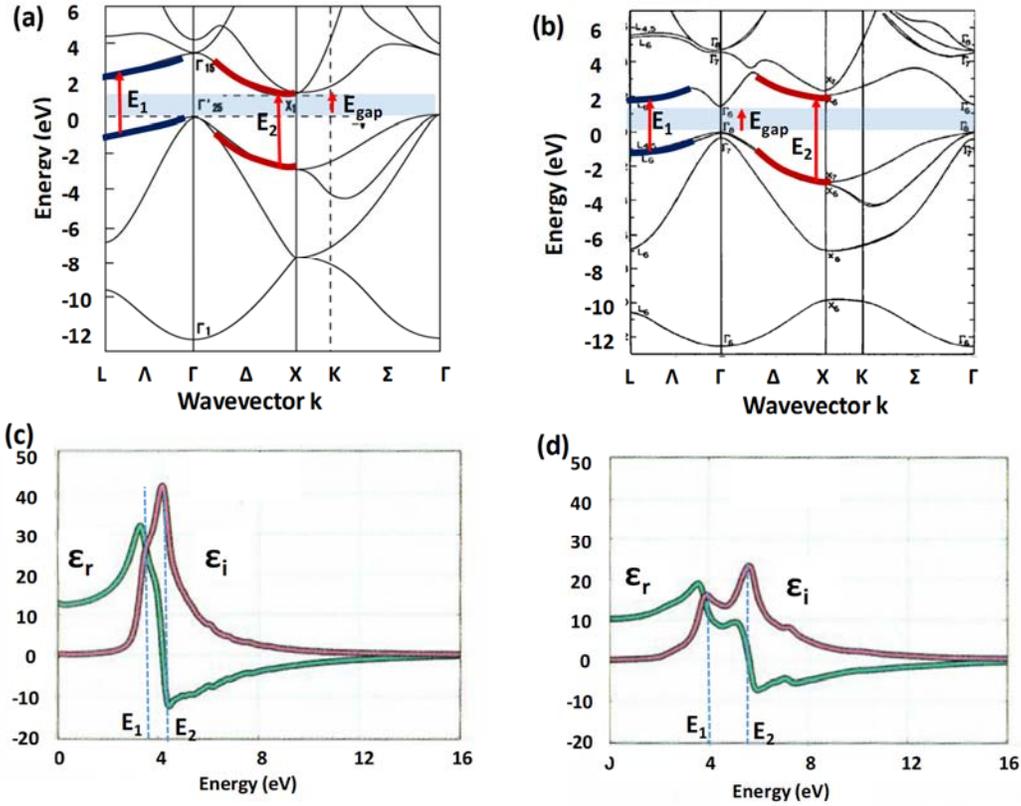

**Figure 3**. Band structures of (a) Si and (b) GaAs with the highlighted states that are the most responsible for the refractive index[38]. Real and imaginary permittivities of (c) Si and (d) GaAs Note that real part of permittivity does shave negative values , but unfortunately only in the high loss region.

Here one should point out the key difference between Lorentz model and actual solid state. In Lorentz model residual absorption far away from resonance decays slowly as $\Delta\omega^2 /(\omega_0 - \omega)^2$ and therefore it is always present. But Lorentz model fails far from the resonance if there are no real states enabling the transition with conservation of energy. In solid state the transitions below the bandgap energy are possible due to either impurities or phonons, and the absorption decreases exponentially as $\exp[-(E_{gap} - \hbar\omega)/E_0]$ - Urbach tail[39] where for high quality single crystal materials $E_0$ is typically a few tens of $meV$. (Alternative sub-bandgap absorption mechanism due to excitonic transitions is discussed further on). Therefore, once the photon energy is roughly $100 meV$ below the bandgap, the absorption becomes negligibly small.

The strategy in search for the high index material is then to identify the materials in which the difference between the effective resonance energy $E_P$ and absorption edge $E_{gap}$ is smallest and obtain the resonantly enhanced index

$$n_{max} \sim \sqrt{1 + \frac{\hbar \omega_p^2}{E_P^2 - E_{gap}^2}} \qquad (13)$$

Now, in this simple picture one can discern why only a very small enhancement of the refractive index below the absorption edge can be observed with most of the covalent materials. With $E_{gap} = 1.1eV \approx E_P/4.3$ for Si and $E_{gap} = 1.35eV \approx E_P/3.8$ for GaAs, one only expect the moderate change of refractive index by $3-4\%$ relative to very long wavelengths as one can indeed observe in Fig.1. Indeed, for most tetrahedral materials the value of $E_P - E_{gap}$ is nearly constant and ranges between 3.5 and 4.5eV[40] indicating that resonant enhancement of refractive index is unattainable in these materials.

- **WHAT CAN BE DONE ABOUT IT?**

So, let us convey in a simple form the conundrum facing us: to obtain large index a *large* number of *large* dipoles have to be *densely* packed in the unit volume of material, which of course contains built-in self-contradiction. As if it were not enough, they have to be packed is such a way that the transition *does not broaden into wide bands*. Let us see what, if anything, can be done about it?

Narrow valence and conduction bands are typically formed by *d*-shells of atoms, rather than *s* and *p* shells of covalent materials. The best known example is familiar $TiO_2$ (rutile, whose band structure is shown in Fig.4a), in which the VB is composed primarily of oxygen p-orbitals while the CB consist of Ti 3d states[41], hence the peak of absorption, as shown in Fig. 4b occurs at a critical point of $4eV$ [33] while the bandedge is around $E_{gap} \sim 3.3eV$. One therefore expects according to (13) a relatively large increase of the refractive index near the resonance. Indeed in bulk rutile refractive index[42] increases from 2.4 at $1.5 \mu m$ to 2.9 at $0.4 \mu m$ - more than a 20% enhancement. In thin film[43] TiO2 refractive index increases from 2.05 to 2.52 – an even larger enhancement near the absorption edge.

One should note that other materials where bonds are at least partially made up by *d*-shells of transition metals, such as LiNbO$_3$ , SrTiO$_3$ [44] and other inorganic perovskites also have large indices and very strong dispersion near the absorption edge[45]. This fact shows that *d*-orbitals may be a key to getting high refractive index at least in a relatively narrow spectral region. In that respect, recently discovered all-inorganic metal halide perovskites may carry a promise[46].

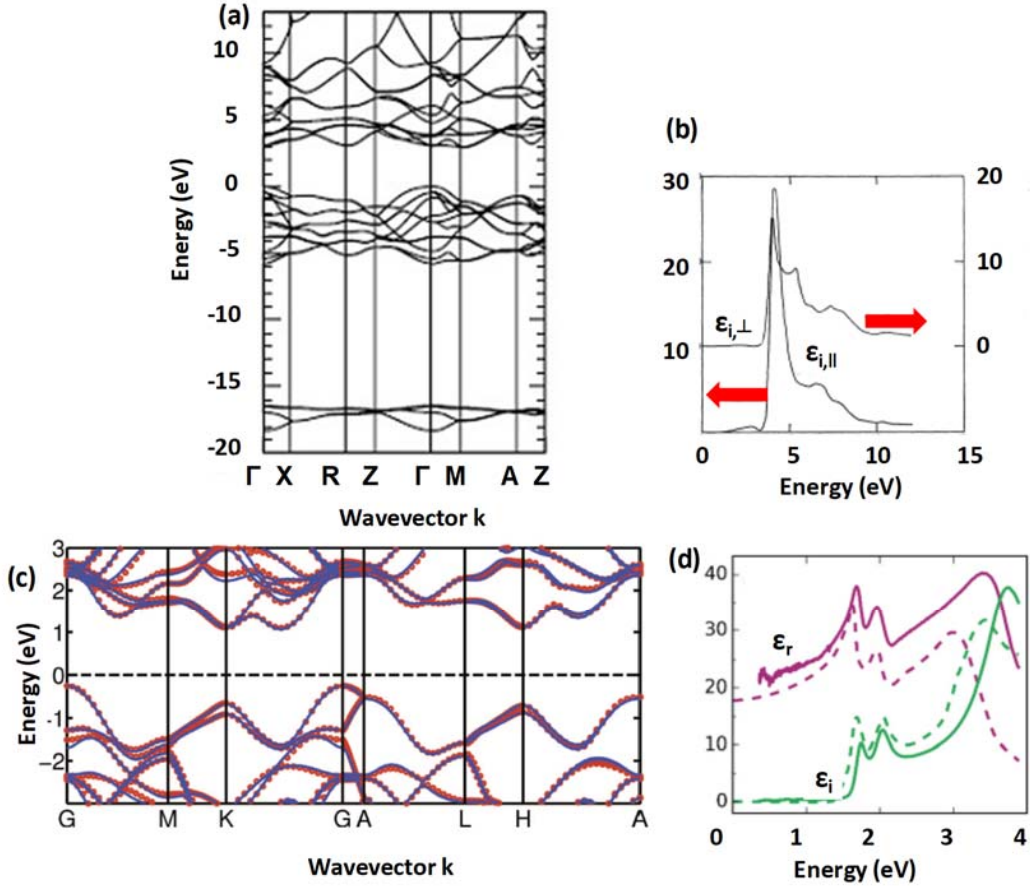

**Figure 4.** (a) Band structure of TiO$_2$ (rutile)[41] (b) Imaginary part of TiO$_2$ permittivity normal and parallel to the axis[33] (c) Band structure of bulk MoS$_2$ [47] (d) Real and imaginary parts of permittivity of MoS$_2$ for monolayer (solid curves ) and bulk (dashed curves)[48]

Another approach to reduction of the width of the band is to consider materials in which some of the bonds have weak, Van-der-Walls (VdW) character, such as transition metal dichalcogenides (TMDC) , WSe$_2$, MoS$_2$ and others[49] . In these materials covalent bonds lying in plane are highly polarizable, while the interlayer coupling is weak and does not contribute much to band broadening. The CB and VB in TMDC's have largely character of *d*-shells of transition metals[47], hence the bands are relatively flat (as seen in Fig.4c) Also, these materials show strong excitonic transitions. With three factors affecting the index: VdW bonding, d-shell character of conduction bands and strong exciton transition **bulk** TMDC's exhibit high indices of 3.6 for W-based and 4.4 for Mo-based [48, 50-52] while being transparent all the way to 600 and 1000nm respectively, which means that these materials outperform $TiO_2$ and Si. Since the performance of focusing elements scales with $n-1$, one can see that $MoS_2$ offers at least 35% improvement relative to Si, while for the PICs where integration density depends on the evanescent wave in the cladding, improvement can be even higher. In fact TMDC application to metasurfaces[53] and PICs[54] have already been implemented . The fact that in TMDC the refractive index is mostly enhanced only for polarization normal to axis does not impede these applications, since PICs typically operate with one (TE in this case) polarization) while in meta lenses the light is polarized normal to the optical axis.

So, the recipe for achieving high index is simply trying to operate in the vicinity of relatively narrow resonance. Such resonances seem to appear stronger in VdW materials due to three factors mentioned above (d shells, VdW bond, exciton), and since only a relatively small fraction of these family has been explored, it is not unfathomable that some material may show refractive index of 5-6 in near IR and maybe 4 through the entire visible range. Since monolayers of VdW materials tend to have wider bandgap than bulk, one may consider heterostructures in which single layers of VdW materials are separated by wide bandgap spacers, like BN[55] – this way a somewhat smaller index can be traded for a wider transmission window.

On a more futuristic level, the writing is on the wall – the way to increase index is to somehow develop a stoichiometric high density arrangement of atoms with strong and narrow transitions. That would require the active atoms to be widely separated, causing reduction in density and hence plasma frequency. But the reduction in the width of the band, determined by the overlap of wavefunctions on different atomic sites would be exponential, and therefore overall the resonant refractive index would be high. Indeed, according to (6) if the atomic spacing is reduced by a factor of 2 from 1.5 Å to 3 Å $\omega_p^2$ gets smaller by a factor of 8, while the bandwidth, according to simple tight binding calculation gets reduced from a few eV to less than a hundred of meV to provide strong resonant enhancement of index. The question is how to keep separation between ions to such large distances, clearly exceeding the bond lengths in any material (the scale of the bod length is dictated by the Bohr radius). Perhaps one can use lattices with under-oxidized alkali-earth metal ions such as $Cs^+$ or $Ba^+$ incorporated *stoichiometrically* into the wide-bandgap lattices of LiF or KCl substituting for alkali metal ions. These ions exhibit very strong and narrow transitions[56-57] in the near UV –blue part of the spectrum that have been used for laser cooling of these ions placed in ions traps. The challenge is of course replacing ion trap by a solid lattice and at this point I can only speculate about feasibility of such arrangement.

One can also, of course, consider metallic artificial structures with high effective refractive index[1, 58]. This concept of "artificial dielectrics"[59] has been known since 1940's[60-61] and has been successfully applied in, for instance, radar antennas operating in the RF and microwave ranges[62] Simply arranging metal spheres into regular lattice one can obtain very high refractive index near localized surface plasmon polariton resonance. Alternatively, layered metal structures force the electric field inside the air gaps which increases effective index[63-64]. These structures work very well from RF to THz[65] where the metal losses are small since the field does not penetrate metal. But in the near IR and visible ranges the losses are enormous[10] and impossible to compensate[66] without reverting to science fiction concepts. The high loss, besides being bad by itself, also reduces the key Q parameter to about 10-20 which makes high index unattainable.

Since even in the absence of all scattering the metal structures with less than 100nm feature size retain high loss due to Landau damping[67], the only way to reduce metal loss is by opening the wide gaps below and above the conduction band. The way to achieve it as shown in [68-69] is either by considering artificial structures with widely separated ions[68] or with(not yet grown) novel VdW material such as $TaS_2$[69], i.e the same natural or composite materials as the ones suggested for high index. Of course, it is entirely expected as presence of strong and narrow resonance portends high permittivity on the low frequency side of the resonance and negative permittivity on the high frequency side (see Fig.3c,d) which would have been very useful had it not been obscured by high absorption. This region of negative permittivity, known as a Reststrahlen band is well known in the IR region where resonances are associated with optical phonons, but unfortunately the wavelength of optical phonon resonances does not get shorter than about $7 \mu m$ in hexagonal BN[70].

It is tantalizing to speculate on whether negative permittivity based on electronic transitions can be observed in the region of low loss. In Ref[71] negative permittivity $\varepsilon \sim -5$ has been measured in dye doped polymer around $\lambda \sim 580 nm$ but the imaginary part of $\varepsilon_i \sim 4$ was too high. Note that very large positive $\varepsilon \sim 16$ has also been measured at longer wavelength of $\lambda \sim 600 nm$. Therefore, the same strategy can be employed to achieve high and negative permittivity combined with low loss, and, as shown in[72-73], for most applications, such as PIC it is preferable to operate in high $\varepsilon$ polaritonic rather than negative $\varepsilon$ Reststrahlen regime..

This brings us to an interesting state of affairs regarding all types of metal-containing optics in visible and near IR ranges, which can be best described as "metamaterial Catch-22"[74-75]. Namely, with the high losses in the existing metals, such contrivances as "superlens" and "hyperlens" do not live up to their lofty promises[14-15, 76-78] If, on the other hand, a really low loss metal (or other negative $\varepsilon$ medium )is developed, then the aforementioned concepts completely lose their appeal, as more or less conventional optics with artificial high index dielectric can provide very high resolution and magnified image in the far field and easily outperform both super and hyper lenses. In other words, with reduction of metal loss the concepts of super and hyper lensing go straight from unrealizable to unnecessary.

- **CONCLUSIONS**

To conclude, this paper has been conceived as less of a review or even a tutorial-like discourse on why the refractive indices are so low, but more as a call for action. It is my belief that even though the basic laws of physics and the values of fundamental constants comprising the Bohr radius leave very little hope for the materials with indices exceeding say $n \geq 10$ in the visible and near IR,, modern methods of growth and fabrication supported by extensive modeling may in the near future lead to relatively modest improvements to perhaps $n \sim 5 - 6$ over a relatively narrow (100nm) band. These modest improvements will, nevertheless, have disproportionally high impact on imaging, integrated photonics and other walks of photonics. In this work, not considering myself a bona fide condensed matter expert by far, I nonetheless have given here a few directions in which the search for these materials may proceed: d-shell transitions, high anisotropy due to VdW bonding, strong excitonic transitions and composites with wide separated metal atoms. This list is definitely far from being complete and I am quite sure that different and probably better pathways will emerge in the future. To achieve the envisioned here results would require a concerted effort of condensed matter theorists, chemists, and material engineers, not to mention photonics practitioners (I have just listed the entire scope of the ACS Photonics target audience). As long as this essay stimulates at least some interest in mounting this effort, its modest goal can be considered accomplished.

- **AUTHOR INFORMATION**

**Corresponding Author**

**Jacob B Khurgin** – Department of ECE, Johns Hopkins University, Baltimore MD 21218 USA, E-mail jakek@jhu.edu

- **ACKNOWLEDEGEMENT**

The author acknowledges support from Air Force Office of Scientific Research (FA9550-16-10362) and DARPA Defense Sciences Office (HR00111720032). Also acknowledged are helpful discussions with Prof. Owen Miller. Finally, encouragement and dedication of Prof. P. Noir and Dr. S. Artois of JHU are as always greatly appreciated.

- **REFERENCES**